# Factorization method for nonlinear evolution equations

Swapan K. Ghosh [1], Debabrata Pal [2], Aparna Saha [3] and Benoy Talukdar∗ [4]

[1] Department of Physics, Darjeeling Govt. College, Darjeeling 734101, India

[2] Bamunara L. C. D. P. Institution, Panowa, Chandipur-Kanpur, Burdwan 713125

[3] Department of Physics, Visva-Bharati University, Santiniketan 731235, India

[4] Nutan Palli, Bolpur 7312304, India

Email : binoy123@bsnl.in, aparna_phyvb@yahoo.co.in

Abstract. The traditional method of factorization can be used to obtain only the particular solutions of the Liénard type ordinary differential equations. We suggest a modification of the approach that can be used to construct general solutions . We first demonstrate the effectiveness of our method by dealing with a solvable form of the modified Emden-type equation and subsequently employ it to obtain the solitary wave solutions of the KdV, mKdV, Rosenau-Hyman (RH) and NLS equations. The solution of the mKdV equation, via the so-called Muira transform , leads to a singular solution of the KdV equation in addition to the well-known soliton solution supported by it. We obtain the solution of the non-integrable RH equation in terms of the Jacobi $sn(u,m)$ function and show that, although robust, it is structurally different from the KdV soliton. We also find the soliton solution of the NLS equation that accounts for the evolution of complex envelopes in an optical medium.

## 1. Introduction

In two interesting publications Rosu and Cornejo-Peréz [1], and Cornejo-Peréz and Rosu [2] found that the factorization of differential operators provides an efficient method to obtain one particular solution of the Liénard type equations written as

$$\ddot{x} + g(x)\dot{x} + F(x) = 0, \; x = x(t) \; . \qquad (1)$$

Here dots over $x$ denote differentiation with respect to time $t$ and, $g(x)$ and $F(x)$ are arbitrary well behaved functions of $x$. Writing (1) in the factorized form

$$(D - \phi_2(x))(D - \phi_1(x))x = 0, \; D = \frac{d}{dt} \qquad (2)$$

it is rather straightforward to deduce that

$$g(x) = -(\phi_1 + \phi_2 + \frac{d\phi_1}{dx}x) \qquad (3)$$

and

$$F(x) = \phi_1 \phi_2 x \; . \qquad (4)$$

Since (1) has been factored as (2), a particular solution of the equation can be found from the first-order equation

$$(D - \phi_1(x))x = 0. \qquad (5)$$

In applying the factorization method to problems of physical interest one proceeds by writing $\phi_1$ and $\phi_2$ with some pre-factors so as to satisfy (4). Understandably, if the pre-factor for $\phi_1$ is taken as $a$, that for $\phi_2$ should be $a^{-1}$. The chosen results for $\phi_1$ and $\phi_2$ when substituted in (3) leads to an equation for $a$. One can assume that $a$ is a c-number. Alternatively, one can regard $a$ as a function of $x$ say $b(x)$. It is straightforward to see that (i) $a$ satisfies an algebraic equation while (ii) $b(x)$ satisfies a first-order differential equation. In the following we deal with an exactly solvable nonlinear Liénard type differential equation and demonstrate that using the value of $a$ we get from (5) only a particular solution of the equation. On the other hand, a similar use of $b(x)$ leads to the general solution of the problem.

Our equation of interest is the modified Emden-type equation given by [3]

$$\ddot{x} + 3kx\dot{x} + k^2 x^3 = 0, \quad x = x(t) \qquad (6)$$

where $k$ is a real arbitrary constant. Consistently with (4), we take for (6)

$$\phi_1 = akx \qquad (7a)$$

and

$$\phi_2 = \frac{kx}{a}. \qquad (7b)$$

From (3) and (7) we get

$$2a^2 + 3a + 1 = 0 \qquad (8)$$

giving $a = -1$ and $-\frac{1}{2}$. For $a = -1$, $\phi_1 = -kx$. Using this value of $\phi_1$ in (5) we obtain

$$x(t) = \frac{1}{kt - c_1}, \qquad (9)$$

where $c_1$ is a constant of integration which may be either positive or negative. For $c_1 = 0$, $x(t)$ is singular at $t = 0$. For positive and negative values of $c_1$, $x(t)$ will be singular at $t = \frac{c_1}{k}$ and $t = -\frac{c_1}{k}$. This particular singular solution was also noted by Reyes and Rosu [4]. Note that we would arrive at the solution as given in (9) even if we worked with $a = -\frac{1}{2}$.

For the $x$ dependent pre-factor $b(x)$ we have chosen

$$\phi_1 = b(x) \qquad (10a)$$

and

$$\phi_2 = \frac{k^2 x^2}{b(x)}. \tag{10b}$$

Equations (10a) and (10b) via (3) give

$$xb'(x) + b(x) + 3kx + \frac{k^2 x^2}{b(x)} = 0, \tag{11}$$

where the prime on $b(x)$ denotes differentiation with respect to $x$. Although nonlinear, we could solve the first-order equation in (11) to get

$$b(x) = \frac{1 - k^2 x^2 + \sqrt{1 - k^2 x^2}}{x}. \tag{12}$$

For clarity of presentation, we took the constant of integration as zero in writing (12). Equations (5), (10a) and (12) give

$$x(t) = \frac{2t}{1 + kt^2}. \tag{13}$$

The result in (13) represents the non-singular general solution of (6) obtained by Leach [3] about three decades ago.

The object of the present work is to follow the viewpoint in (ii) and thereby construct general solutions of a number of nonlinear evolution equations (NLEEs) which play a role in several applicative contexts ranging from soliton formation in shallow water waves to propagation of signals through optical fibers [6]. The equations of our interest are the well-known KdV, mKdV (modified KdV), Rosenau-Hyman (RH) and nonlinear Schrödinger (NLS) equations [5]. These equations are partial differential equations in (1+1) dimensions. In section 2 we convert them to ordinary differential equations (ODEs) using transformation to the traveling variable and then implement the above modification of the original factorization method [1,2] to construct their solutions. We also provide some physical interpretation for the results obtained. We devote section 3 to summarize our outlook on the present work and conclude by noting that an obvious virtue of our method is its simplicity and provides an uncomplicated supplement for other sophisticated techniques [7,8] often used to solve NLEEs.

## 2. Traveling-wave solutions

In the standard notation the KdV, mKdV, RH and NLS equations are written as

$$v_t - 6vv_x + v_{xxx} = 0, \quad v = v(x,t), \tag{14}$$

$$v_t - 6v^2 v_x + v_{xxx} = 0, \tag{15}$$

$$v_t + 3v^2 v_x + 6v_x v_{xx} + 2vv_{xxx} = 0 \tag{16}$$

and

$$iu_x + \frac{1}{2} u_{tt} + |u|^2 u = 0, \quad u = u(x,t). \tag{17}$$

respectively.

In the above equations $x$ and $t$ stand for the space and time variables. The subscripts on $v$ and $u$ denote appropriate partial derivatives. The $v$ fields are real while $u$ refers to a complex field. We shall now construct their exact solutions.

(i) KdV equation

Introducing the traveling variable $\varsigma = x - ct$ with $c$, the velocity of the wave and defining $v(x,t) = f(\varsigma)$ one can convert (14) to an ordinary differential equation given by

$$f''' - 6ff' - cf' = 0,  \qquad (18)$$

where primes denote differentiation with respect to $\varsigma$. In particular, $f''' = \dfrac{d^3 f}{d\varsigma^3}$. Equation (18) can be integrated to get

$$f'' - cf - 3f^2 = 0. \qquad (19)$$

In writing (19) we imposed the boundary conditions $f''(\varsigma) = f'(\varsigma) = f(\varsigma) = 0$ at $\varsigma = \pm\infty$. The Liénard type equation in (19) with the independent variable $\varsigma$ and dependent variable $f$ can be solved by writing

$$\phi_1 = b(f) \quad \text{and} \quad \phi_2 = -\frac{c + 3f}{b(f)}. \qquad (20)$$

Equation (20) when substituted in (3) written in terms of $f$ leads to the first-order linear ordinary differential equation

$$\frac{d\psi}{df} + \frac{2}{f}\psi = \frac{2(c + 3f)}{f}, \psi = b^2(f). \qquad (21)$$

This equation can be integrated to get

$$b(f) = \sqrt{c + 2f}. \qquad (22)$$

The value of $b$ in conjunction with

$$\frac{df}{d\varsigma} - b(f)f = 0 \qquad (23)$$

gives the well known solution

$$v(x,t) = \frac{c}{2}\operatorname{sech}^2 \sqrt{c}(x - ct) \qquad (24)$$

for the KdV equation. The wave represented by (24) propagates undistorted and has been given the name soliton [9]. For simple physical realization of undistorted wave propagation we note that the linear term $(v_{xxx})$ and nonlinear term $(vv_x)$ in (14) have opposite effects on the evolution of $v(x,t)$; the first term causes dispersion of the wave while the second one leads to steepening. A balance between these two effects produces a traveling wave of constant shape. Physically, solitons are localized waves that propagate without change of their properties. These localized waves are stable against mutual collision. The first one is a solitary wave condition known in hydrodynamics. The second one means that the waves have particle like properties.

(ii) mKdV equation

In the traveling variable the ordinary differential equation corresponding to (15) is given by

$$f''' - 6f^2 f' - cf' = 0. \tag{25}$$

Integrating (25) with respect to $\varsigma$ and imposing appropriate boundary conditions we get

$$f'' - f(2f^2 + c) = 0. \tag{26}$$

Equation (26) suggests that we should choose $\phi_1$ and $\phi_2$ as

$$\phi_1 = b(f) \text{ and } \phi_2 = -\frac{2f^2 + c}{b(f)} \tag{27}$$

which when used in (3) lead to the first order linear differential equation

$$\frac{d\psi}{df} + \frac{2}{f}\psi - \frac{2}{f}(2f^2 + c) = 0, \ \psi = b^2(f). \tag{28}$$

From the solution of (28) we have

$$b(f) = \sqrt{f^2 + c}. \tag{29}$$

Substituting this value of $b(f)$ in (23), the resulting differential equation can be solved to get

$$v(x,t) = v_m(x,t) = -\frac{4\sqrt{c}e^{\sqrt{c}(x-ct)}}{4e^{2\sqrt{c}(x-ct)} - 1}. \tag{30}$$

We used the subscript $m$ on $v(x,t)$ merely to indicate that (30) represents a solution of the mKdV equation. Clearly, it is a singular solution. A similar solution was constructed by Rasinariu, Sukhatme and Khare [10] by using Matveev's generalized Wronskian formula of the KdV equation [11].

The solutions of the KdV and mKdV equations are related by the nonlinear transformation [12]

$$v_{1,2} = v_m \pm \frac{\partial v_m}{\partial x} \tag{31}$$

called the Miura transformation. From (30) and (31) we obtain two other solutions of the KdV equation as

$$v_1(x,t) = \frac{-4c}{(2e^{\sqrt{c}(x-ct)} - e^{-\sqrt{c}(x-ct)})^2} \tag{32}$$

and

$$v_2(x,t) = \frac{-4c}{(2e^{\sqrt{c}(x-ct)} + e^{-\sqrt{c}(x-ct)})^2}. \tag{33}$$

The result in (32) represents a singular solution of the KdV equation and cannot be found by standard methods used for solving it [10]. The negative of the non-singular second solution, namely, $-v_2(x,t)$ closely resemble the soliton solution in (24).

   (iii)   RH equation

Looking at (14) and (15) we see that the dispersive term ($v_{xxx}$) in the KdV and mKdV equations are linear. Such equations are often referred to as quasi-linear. But the dispersive term ($vv_{xxx}$) of the RH equation in (16) is nonlinear. It is thus fully nonlinear. The fully nonlinear equations can have solitary wave solutions with compact support. These solutions are called compactons [13]. It is well known that solitons are characterized by exponential tails. As opposed to this compactons vanish identically outside a finite range but are robust within their range of existence. The soliton was first observed in shallow water waves [14] and play a role in many problems in hydrodynamics. The compactons are useful in the study of pattern formation in liquids.

For the RH equation, the ordinary differential equation in the traveling can be found as

$$2ff''' + 6ff'' + 3f^2 f' - cf' = 0 \tag{34}$$

which on integration yields

$$ff'' + f'^2 + \frac{1}{2}f(f^2 - c) = 0. \tag{35}$$

Due to the presence of the term $f'^2$, (35) is not a Liénard type equation and provides some difficulty to write the equation in the factorable form. However, we express it as

$$(\frac{d}{d\varsigma} - \phi_2)(f\frac{d}{d\varsigma} - \phi_1)f = 0. \tag{36}$$

In the expanded form (36) reads

$$ff'' + f'^2 - (\phi_1 + f\phi_2 + f\frac{d\phi_1}{df})f' + \phi_1\phi_2 f = 0. \tag{37}$$

For (35) and (37) to be identically equal one requires the coefficient of $f'$ in the latter equation to vanish. This gives

$$\phi_1 + f\phi_2 + f\frac{d\phi_1}{df} = 0 \tag{38}$$

and at the same time allows us to assume

$$\phi_1 = b(f) \text{ and } \phi_2 = \frac{f^2 - c}{2b(f)} \tag{39}$$

From (38) and (39) we obtain

$$\frac{d\psi}{df} + \frac{2}{f}\psi + f^2 - c = 0. \tag{40}$$

Solving (40) we get

$$b(f) = \sqrt{5cf - 3f^3}. \tag{41}$$

With this value of $b(f)$ we solved

$$(f\frac{d}{d\xi} - \phi_1)f = 0 \tag{42}$$

to obtain the solution of the RH equation in the form

$$v(x,t) = \sqrt{\frac{5}{3}}[2sn^2(\frac{i(x-ct)}{2(15)^{\frac{1}{4}}},2) - 1], \tag{43}$$

where $sn(w,k)$ stands for a Jacobi elliptic function [15]. It may be an interesting curiosity to compare the plot of the soliton solution of the KdV equation with that (compacton) of the RH equation.

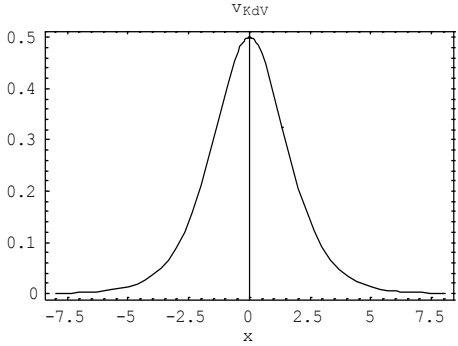 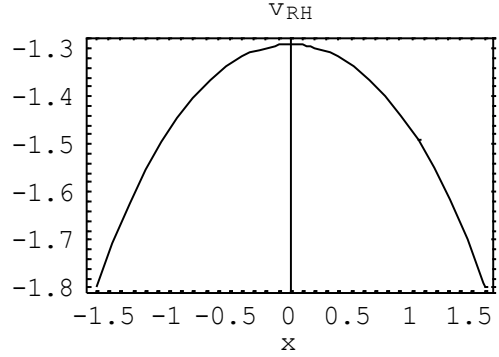

Fig.1. KdV soliton as function of $x$.    Fig.2. RH compacton as a function $x$.

`In Figs 1 and 2 we plot the soliton and compacton solutions of (14) and (16) as a function of $x$ for $t = 0$ and $c = 1$. It is clear from these figures that the KdV soliton is exponentially localized while the compacton solution of the RH vanish identically near $x \approx \pm 1.48$. Physically, the observed difference between the soliton and compacton structures may be attributed to variation in the dispersive effects caused by the last two terms in (14) and (16).

(iv)    NLS equation

Since the discovery of solitons in water waves [14] the solitary wave phenomenon attracted attention of scientists working in different fields. The appearance and applications of solitons are too numerous to be listed. Remarkable among them is the crucial impact of the NLS solitons in fiber optics. Here, solitons are used as information carrier. In the optical context, the term soliton is used to refer to any optical field that does not change during the propagation because of a balance between nonlinear and dispersive effects in the medium. The nonlinear term in the governing equation is a consequence of the Kerr effect in the medium. The Kerr effect is a phenomenon where an applied field induces a change in the refractive index of the material [16].

We have written the NLS equation in the optical notation with the space and time variables interchanged as compared to notations used in (14) – (16). We are interested in solutions of (17) in the form

$$u(x,t) = f(t - \alpha x)\exp(-i\omega t + ikx) \tag{44}$$

These solutions are the simplest in a large family of quasi-periodic solutions to the nonlinear Schrödinger equation [17]. In terms of the traveling variable

$$\tau = t - \alpha x \tag{45}$$

(44) reads

$$u(x,\tau) = f(\tau)\exp(-i\omega\tau - i\omega\alpha x + ikx). \tag{46}$$

Here $\alpha$, $k$ and $\omega$ denote the reciprocal velocity, wave number and angular frequency respectively. From (17), (45) and (46) we obtain the complex equation

$$\frac{1}{2}\alpha^2 f'' + (\omega - \frac{1}{2}(k-\alpha\omega)^2)f + f^3 + i(1-\alpha(k-\alpha\omega))f' = 0. \tag{47}$$

Equating the real and imaginary of (47) we have

$$\frac{1}{2}\alpha^2 f'' + (\omega - \frac{1}{2}(k-\alpha\omega)^2)f + f^3 = 0 \tag{48}$$

and

$$(1-\alpha(k-\alpha\omega))f' = 0. \tag{49}$$

From (49) we get a constraint on the physically admissible values $\alpha$, $k$ and $\omega$ such that

$$\alpha(k-\alpha\omega) = 1. \tag{50}$$

In view of (50), the dynamical equation (48) can be written as

$$f'' + [\frac{1}{\alpha^2}(2\omega - \frac{1}{\alpha^2}) + \frac{2}{\alpha^2}f^2]f = 0. \tag{51}$$

For (51) we choose

$$\phi_1 = b(f) \text{ and } \phi_2 = [\frac{1}{\alpha^2}(2\omega - \frac{1}{\alpha^2}) + \frac{2}{\alpha^2}f^2]/b(f). \tag{52}$$

As in the earlier cases we can find an equation for $b(f)$ and solve it to get

$$b(f) = \pm\frac{\sqrt{1-2\alpha^2\omega^2 - \alpha^2 f^2}}{\alpha^2}. \tag{53}$$

Here we have taken the constant of integration as zero. Using $\phi_1 = b(f)$ with the negative sign before the radical we solved an equation similar to that in (23) to write

$$f(\tau) = \frac{4\sqrt{1-2\alpha^2\omega}\, e^{\frac{\tau}{\alpha^2}\sqrt{1-2\alpha^2\omega}}}{4\alpha^2 + e^{2\frac{\tau}{\alpha^2}\sqrt{1-2\alpha^2\omega}}} \tag{54}$$

which for $\alpha = \frac{1}{2}$ gives the well known result

$$|u(x,\tau)| = A \sec h(A\tau), \quad A = 4\sqrt{1 - \frac{\omega}{2}} \ . \qquad (55)$$

**3. Concluding remarks**

In this work we constructed the traveling wave solutions of three quasi-linear ( KdV, mKdV and NLS) and one fully nonlinear (RH) evolution equations. The quasi-linear equations support exponentially localized  soliton solution . On the other hand, the fully nonlinear equation was found to have a similar robust solution with compact support . The modern theory of solitons was born in 1967 when Gardner, Greene, Kruskal and Muira [7] related the Cauchy initial value problem for the KdV equation to the inverse scattering problem [18] for a one-dimensional linear Schrödinger equation . In 1972 Zakharov and Shabat [19] showed that the method of inverse scattering transform can also be adapted to obtain the soliton solution of the NLS equation. Almost simultaneously with the work of Zakharov and Shabat, Wadati [20] derived a similar approach for the mKdV equation. In addition to inverse spectral methods, there are also some special methods to solve nonlinear evolution equations. Remarkable among them is the Hirota's bi-linear method [21]. The idea behind Hirota's method was to transform the original evolution equation into new variables so that in these variables the multi-soliton solutions can be found in a rather simple form. These approaches are technical enough to be followed by a wide variety of physicists and  mathematicians.  For example, the use of the inverse spectral method requires a good background of the theories of integral equations in addition to having mastery over quantum mechanics. In this context we note that  during the last few decades the necessary mathematical tool of solving NLEEs have become more sophisticated [22]. Since solitons appear in a variety of physical contexts it is  of considerable to have in the literature new  uncomplicated methods to solve nonlinear evolution equatiom.

   Keeping the above in view we provided in this paper a direct method for solving nonlinear evolution equations by a judicious use of the so-called factorization method as originally developed in  refs. 1 and 2. The use of our method requires only the knowledge of elementary calculus. Our solution of the mKdV equation via Muira tranform led to a singular solution  of the KdV equation in addition to the usual soliton solution supported by the latter. In ref.10 such singular solutions was obtained by a fairly complicated procedure. As opposed to the case of soliton equations, namely, the KdV, mKdV and NLS equations, the application of the factorization method to the fully nonlinear RH equation does not lead to  Liénard type equations. Consequently, we had to introduce a different factorization scheme. This can be seen by comparing (2) and (36). Also we point out that the result in (43) giving an expression for the compacton is new to the best of our knowledge.

**3.Reference**